\documentstyle[12pt,aasms4,psfig]{article}
\begin{document} 
\title{COMPARISON OF ANALYTICAL MASS FUNCTIONS WITH 
NUMERICAL SIMULATIONS}
\author{Jounghun Lee$^{1,2}$ and Sergei F. Shandarin$^{1,3}$} 
\affil{$^1$Department of Physics and Astronomy,
University of Kansas, Lawrence, KS 66045 \\ 
$^2$ Institute of Astronomy and Astrophysics, Academia Sinica, 
Nankang, Taipei, Taiwan \\ 
$^3$ Theoretical Astrophysics Center, Juliane Maries Vej 30,
DK-2100 Copenhagen, Denmark\\
taiji, sergei@kusmos.phsx.ukans.edu}

\begin{abstract} 

We present numerical testing results of our mass function derived in our 
previous paper, and compare the testing results with those of the popular 
Press-Schechter (PS) mass function.  Two fiducial models are 
considered for the test: the scale-free power-law spectra  
$P(k) \propto k^{n}$ 
with spectral indices $n=-1, 0$ and the standard cold dark matter 
(SCDM) model with $\Omega =1$ and $h=0.5$.  
For the power-law models,  we use numerical data averaged 
over several different output times: ten output times from two
N-body realizations for the $n=-1$ power-law model; 
four outputs from one realization for the $n=0$ model.     
While for the SCDM model,
we consider four outputs separately at redshifts $z = 0, 0.43, 
1.14$ and $1.86$ from one large N-body simulations.  
The comparison results show that our mass function
fits the numerical data in a much improved way over the PS one.  
Thus,  we expect that our mass function can be a viable alternative
of the PS mass function in applications to various areas. 

\keywords{cosmology: theory --- large-scale structure 
of universe}

\end{abstract}

\section{INTRODUCTION}

One of the most fundamental problems in cosmology is to understand 
the formation and evolution of the large-scale structure in the universe 
such as galaxies, groups and clusters of galaxies, etc.  
In order to understand the large-scale structure,  however, 
it is highly desirable to have an analytical framework 
within which theoretical predictions for structure formation can be made. 
The cosmological mass function, $n(M)$  
[$n(M)dM$: defined as the comoving number density of gravitationally 
bound structures -- dark halos with mass $M$] 
provides this analytical tool
since different candidate models for structure formation 
predict different number densities of dark halos. 

Press \& Schechter (1974, hereafter PS) developed for the first time 
an analytic formalism to evaluate the mass function.  
Finding a mass function requires both dynamics and statistics.  
Dynamically PS adopted the top-hat spherical model, according to which
the collapse condition for forming dark halos is determined purely
by its local average density. 
Statistically PS assumed that the initial density field is Gaussian, 
and selected dark halos from the peaks of the linear Gaussian density
field. 

The practical success of the PS mass function 
(e.g., \cite{efs-etal88}; \cite{la-co94}) and the absence of 
viable alternatives led many authors to use it 
routinely in the last decade(e.g., \cite{col-kai88}; 
\cite{whi-fre91}; \cite{kau-etal93}; \cite{fan-etal97}). 
  
But, recently high resolution N-body simulations have shown 
the limitation of the PS theory:  
First, it has been clearly shown by several N-body simulations that 
the true gravitational collapse process must be nonspherical 
(e.g., \cite{sh-etal95}; \cite{kuh-etal96}), 
indicating the weakness of the PS dynamical background.      
Second, N-body simulations also showed that
peaks of the linear density field are ``poorly'' correlated with 
the final spatial locations of dark halos (e.g., \cite{kat-etal93}),
although the PS formalism assumes that dark halos form in the 
peaks of the density field. 
Third, recent high-resolution numerical tests have detected that 
experimental results are flatter than the standard PS mass 
function in shape 
(e.g., Governato et al. 1998; Tormen 1998 and references therein
; \cite{she-tor99}).

Currently various attempts have been made to find a better-fitting 
mass function. One approach to better mass functions 
has been focused on finding phenomenological fitting parameters, 
keeping the original PS formula unchanged; 
for example, regarding the density threshold as a function of 
redshift, etc   
(see Governato et al. 1998; \cite{she-tor99}). 
Another approach has been on improving the dynamics of the PS formalism 
by implementing anisotropic collapse conditions 
(e.g., \cite{mo95}; Audit, Teyssier, \& Alimi 1997). 
A full analytical alternative of the PS mass function using  
this approach has been found by \cite{lee-sha98}, in the frame of the 
Zel'dovich approximation.  This approach goes along with the assumption 
that dark halos correspond to the third axis collapse.   
Indeed the mass 
function we found in the previous paper is both a dynamically and 
statistically improved version of the PS one. First, it is based on
a more realistic {\it nonspherical} dynamical model. Second, the 
underlying statistical assumption that dark halos form in the 
local maxima of the smallest eigenvalue of the deformation 
tensor, $\lambda_{3}$ (see $\S 2$) is in general 
agreement with the N-body results performed by \cite{sha-kly84}. 
Third, and most importantly our mass function was shown to have desired 
properties like a lower peak and more high-mass halos.  That is,  
our mass function is flatter than the PS one. 

In this Letter we present numerical testing results of our mass 
function for the case of two fiducial models: the scale-free
power-law spectra with spectral indices $n=-1, 0$, and the standard 
cold dark matter (SCDM) model with $\Omega =1$ and $h=0.5$.   
In  $\S 2$ we briefly summarize the analytic mass function theories 
for the readers convenience . 
In $\S 3$ we explain the N-body simulations used to produce 
the numerical mass functions, and compare the analytical mass functions
with the numerical results.  
In $\S 4$ we draw a final conclusion. 

\section{SUMMARY OF MASS FUNCTION THEORIES}

The PS theory assumes that dark halos of mass $M$ form
hierarchically in the regions where the linear Gaussian density
field $\delta \equiv (\rho - \bar{\rho})/\bar{\rho}$ 
($\bar{\rho}$: mean density) filtered on mass scale $M$ reaches 
its threshold value $\delta_{c}$ for collapse: 
\begin{equation}
n_{PS}(M) = \sqrt{\frac{2}{\pi}}\frac{\bar{\rho}}{M^2}
\Bigg{|}\frac{d\ln\sigma}{d\ln M}\Bigg{|}
\frac{\delta_{c}}{\sigma}\exp\bigg{[}-\frac{\delta_{c}^2}
{2\sigma^2}\bigg{]}.  
\end{equation} 

The density threshold $\delta_{c}$ for a flat universe  
is originally given by the spherical top-hat model: 
$\delta_{c} \approx 1.69$ (e.g., \cite{pee93}).  
But in many numerical tests 
it has been detected that lowered $\delta_{c}$ (roughly $1.5$)  
gives a better fit in the high-mass section 
(e.g., \cite{efs-ree88}; \cite{car-cou89}; \cite{kly-etal95}; 
\cite{bo-my96}).  It is also worth mentioning that 
any PS-like formalism is least reliable in the low-mass section 
(\cite{mo95}). 
This numerical detection can be understood in the following 
dynamical argument:  Although the top-hat spherical model 
predicts that the gravitational collapse to ``infinite'' density 
occurs when the density reaches $\delta_{c}\approx 1.69$, halos 
in realistic case can form earlier by a rapid virialization 
process due to the growth of small-scale inhomogeneities (\cite{sha-etal99}). 

On the other hand, according to our approach (Lee \& Shandarin 1998),  
dark halos of mass $M$ form from the Lagrangian regions 
where the lowest eigenvalue $\lambda_{3}$ 
($\lambda_3<\lambda_2<\lambda_1, ~~~ 
\delta =\lambda_1+\lambda_2+\lambda_3$)
of the deformation tensor $d_{ij}$ (defined as the second derivative 
of the perturbation potential  $\Psi$ such that 
$d_{ij}=\partial^2 \Psi /\partial q_i\partial q_j$, $q_i$ is the Lagrangian 
coordinate) reaches its threshold $\lambda_{3c}$ for collapse 
on the scale $M$:
\begin{eqnarray}
n_{LS}(M) &=& \frac{25\sqrt{10}}{2\sqrt{\pi}}
\frac{\bar{\rho}}{M^2}\Bigg{|}
\frac{d\ln\sigma}{d\ln M}\Bigg{|}\frac{\lambda_{3c}}{\sigma}
\Bigg{\{}
\Big{(}\frac{5\lambda_{3c}^2}{3\sigma^2}-\frac{1}{12}\Big{)}
\exp\Big{(}-\frac{5\lambda_{3c}^2}{2\sigma^2}\Big{)}
{\rm erfc}\Big{(}\sqrt{2}\frac{\lambda_{3c}}{\sigma}\Big{)}
\nonumber \\
&& +\frac{\sqrt{6}}{8}
\exp\Big{(}-\frac{15\lambda_{3c}^2}{4\sigma^2}\Big{)}
{\rm erfc}\Big{(}\frac{\sqrt{3}\lambda_{3c}}{2\sigma}\Big{)}
-\frac{5\sqrt{2\pi}\lambda_{3c}}{6\pi\sigma}
\exp\Big{(}-\frac{9\lambda_{3c}^2}{2\sigma^2}\Big{)}
\Bigg{\}}.  
\end{eqnarray}

In the original derivation of our mass function, the threshold 
$\lambda_{3c}$ for collapse has been empirically chosen to be $0.37$.  
A similar logic used to give a dynamical explanation to 
the lowered $\delta_{c}$ of the PS formalism applies here. 
Although a simple extrapolation of the Zel'dovich approximation to 
nonlinear regime predicts that the formation of dark halos 
corresponding to the third axis collapse  
occurs at $\lambda_{3c} = 1$, the first and 
the second axis collapse speed up the formation of halos, which 
would result in a lowered $\lambda_{3c}$ (see also \cite{au-etal97}).    
 
In $\S$ 3, we witness that our mass function with 
the original suggested value of $\lambda_{3} = 0.37$ 
does agree with the numerical data quite well.
  
\section{NUMERICAL vs. ANALYTICAL MASS FUNCTIONS} 

\subsection{Comparison for Scale-Free Model}

The N-body simulations of a flat matter-dominated universe for power-law 
spectra $P(k) \propto k^n$ with spectral indices $n = -1$ and $0$  
were run by \cite{whi94} using a Particle-Particle-Particle-Mesh code
with $100^3$ particles in a $256^3$ grid with periodic boundary conditions.  

Tormen (1998, 1999)  
identified dark halos from the N-body simulations using 
a standard halo finder  -- the friends-of-friends algorithm 
with a linking length $0.2$ [hereafter FOF (0.2)]. 
Numerical data for the $n = -1$ power-law model 
were obtained for 10 different output times 
coming from two N-body realizations, and then a final $n=-1$  
numerical mass function was obtained by taking an average over 
the 10 output values.  
While for the $n = 0$ model 4 outputs from one N-body 
realization were averaged to produce a final numerical mass function.   
For a detailed description of the simulations, see \cite{tor-etal97}.   
Here we use the final average numerical mass functions for comparison data. 

For the power-law spectra, 
the mass variance is given by the following simple form:
\begin{equation}
\sigma^{2}(M) = \Bigg{(}\frac{M}{M_{0}}\Bigg{)}^{-(n+3)/3} ,
\end{equation}     
where $M_{0}$ is the characteristic mass scale 
\footnote{In  \cite{lee-sha98}, the characteristic mass was notated 
by $M_{*}$.  But here we use $M_{*}$ to notate a slightly 
different mass scale. Readers should not be confused about this 
different  notation of the characteristic mass.}   
defined by $\sigma(M_{0}) = 1$.   It is sometimes useful to define 
a filter-depending nonlinear mass  scale $M_{*}$ related to $M_{0}$ by 
$M_{*} \equiv M_{0}(\delta_{c})^{-6/(n+3)}$ for a dimensionless 
rescaled mass variable $M/M_*$ such that $\sigma(M_*) = \delta_c$ 
(see \cite{la-co94}).   
    
Figure 1 plots the fraction of mass in halos with mass $M$,  
$dF/d\ln M = (M^{2}/\bar{\rho})n(M)$: 
our mass function with $\lambda_{3c} = 0.37$ (solid line) against 
the averaged numerical data with Poissonian error bars, 
and the PS mass function with $\delta_{c} = 1.69$ and $1.5$ 
(dashed and dotted lines respectively) as well.  The upper panel 
corresponds to the $n=-1$ power-law model while the lower panel to 
the $n=0$ model.  
As one can see, our mass function fits the numerical data much better 
than the PS ones for the $n=-1$ model in the high-mass section 
($M > M_{*}$).
In fact \cite{tor98} also used the spherical overdensity 
algorithm [SO (178)] as another halo finder, and showed that the numerical 
mass functions from FOF (0.2) and SO (178) are almost identical. 
We  compared the analytical mass functions with his numerical data  
obtained from SO (178) and also found similar results.    
Whereas for the $n=0$ model, neither of our mass function and the PS 
one fits the numerical data well in the high-mass section.  
Yet in the low-mass section $(M < M_{*})$ our mass function fits   
slightly better than the PS one for this case.  

\subsection{Comparison for SCDM model}

\cite{gov-etal98} provided halo catalogs produced from  
one large N-body realization (comoving box size of 
$500h^{-1}{\rm Mpc}$, $47$ million particles on a $360^{3}$ grid)
of SCDM model with $\Omega = 1, h = 0.5$ for four different epochs: 
$z = 0$, $0.43$, $1.14$ and $1.86$ which are respectively 
normalized by 
$\sigma(8h^{-1}{\rm Mpc}) = 1.0$, $0.7$, $0.467$ and $0.35$. 
They adopted the transfer function given by \cite{bar-etal86} and 
also used the FOF (0.2) halo finder. For a detailed description 
of the simulations, see \cite{gov-etal98}.  
 
We derived the numerical mass functions from the catalogs 
by directly counting the number densities of halos in logarithmic 
scale for each epoch. 
In accordance with \cite{gov-etal98},  we consider halos more 
than $64$ particles (corresponding to $ M > 10^{14}M_{\odot}$) 
in order to avoid small-number effects of the N-body simulations.  
Figure 2 shows the comparison results.  
Our mass function with $\lambda_{3c} = 0.37$ (solid line) fits the 
numerical data much better than the PS ones with 
$\delta_{c} = 1.69$ and $1.5$ (dashed and dotted lines respectively) 
for all chosen epochs.

\section{CONCLUSION}

We have numerically tested an analytical mass function recently 
derived by \cite{lee-sha98}, and compared the results with that of  
the standard Press-Schechter one.  

Our mass function is not just a phenomenologically obtained 
fitting formula but a new analytic formula derived through 
modification of the PS theory using a nonspherical dynamical model.   
It is based on the Zel'dovich approximation taking 
into account the nonspherical nature of real gravitational collapse
process, while the PS mass function  
is based on the top-hat spherical model.    
Consequently our mass function is characterized by 
the threshold value of the smallest eigenvalue of the deformation 
tensor, $\lambda_{3c}$ while the PS one by the density threshold, 
$\delta_{c}$. 
 
We have shown that in the power-law model  
with spectral index $n=-1$ and the four different epochs of SCDM  
\footnote{At four different epochs of the SCDM model we effectively 
probe the dependence of the fit to the slope of the initial spectrum.} 
our mass function with $\lambda_{3c} = 0.37$ is  
significantly better than the PS one with $\delta_{c} = 1.69 - 1.5$.    
It fits the numerical data well especially in the high-mass section 
(corresponding to groups and clusters of galaxies) for these two 
models.  
Furthermore it is worth noting that in the testing results for SCDM model  
our mass function agrees with the data well with a consistent threshold 
value of $0.37$ at all chosen redshifts. 
 
On the contrary, the testing result for the $n=0$ power-law model has  
shown that there are considerable discrepancies between the analytical 
mass functions (both of our mass function and the PS one) and the 
numerical data in the high-mass section.  The discrepancies with 
theory for the $n=0$ model, however, have been already detected 
(\cite{la-co94}).      
Yet in the low-mass section our mass function fits the data slightly 
better for this case.      

Although we have tested our mass function only for two different models, 
given the promising testing results of our mass function demonstrated here, 
we conclude that it will provide a more accurate 
analytical tool to study structure formation.  Further testings of the   
new mass function are obviously very desirable and will be reported 
in the following publications.  

\acknowledgments

We are very grateful to Giuseppe Tormen who provided the $n=-1,0$ numerical 
mass functions, and also for serving as a referee 
in helping to improve the original manuscript.  
We are also grateful to Fabio Governato's  
SCDM halo catalogs and useful comments.
We acknowledge the support of EPSCoR 1998 grant. 
S. Shandarin also acknowledges the support of GRF 99 grant 
from the University of Kansas and the TAC Copenhagen.

\begin{figure}[tb]
\psfig{figure=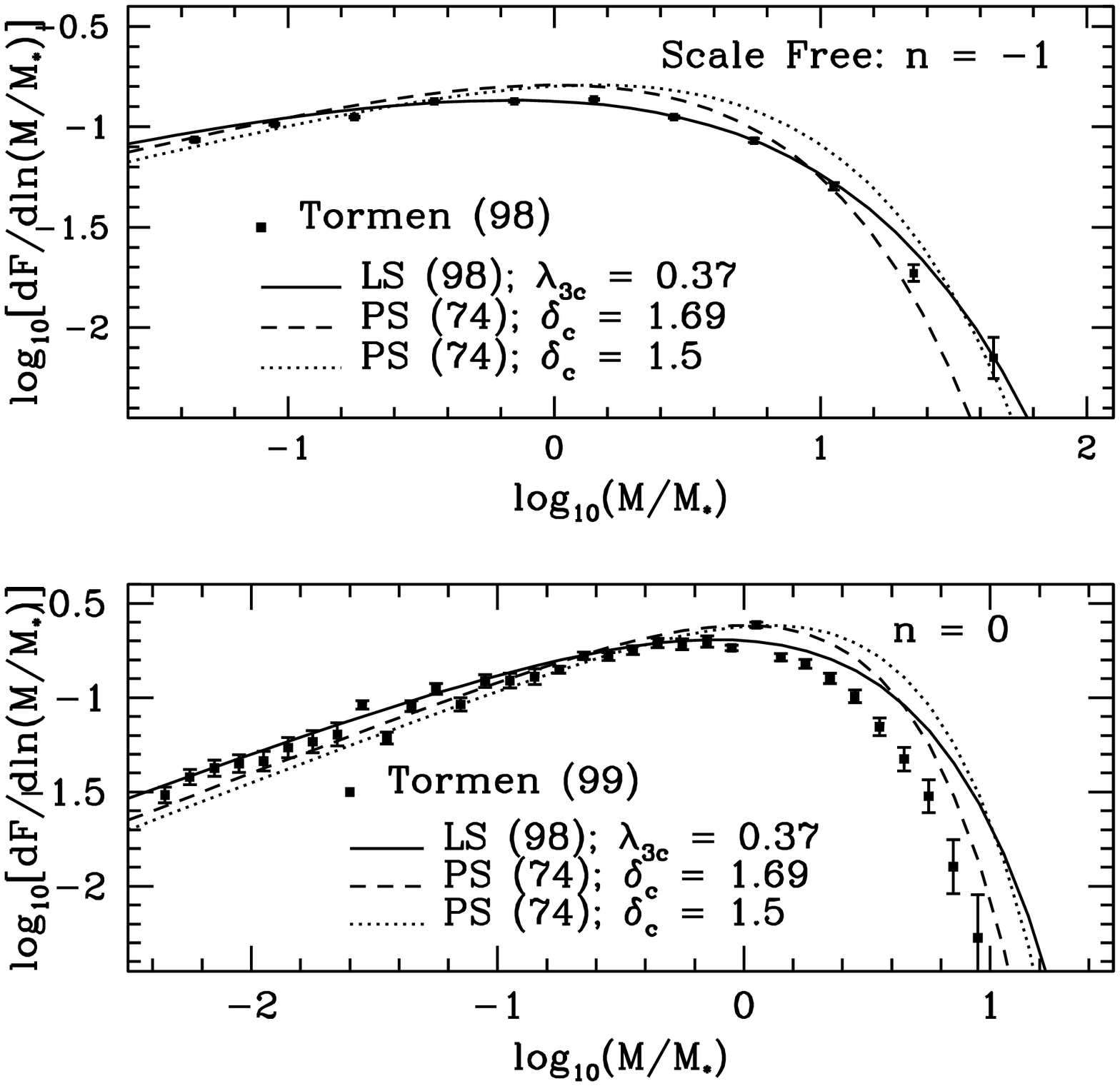,height=14.5cm,width=15.5cm}
\caption[fig1.ps]{The square dots represent the numerical mass
function with poissonian error bars. The solid line is the LS mass 
function with $\lambda_{3c} = 0.37$ while 
 the dashed, the dotted lines are the PS mass functions with 
$\delta_{c} = 1.69, 1.5$ respectively.  The upper and the lower 
panels correspond to the $n=-1$ and the $n=0$ power-law models respectively. 
See also the top left panel of Fig.2 in \cite{tor98}. }
\end{figure}
\begin{figure}[tb]
\psfig{figure=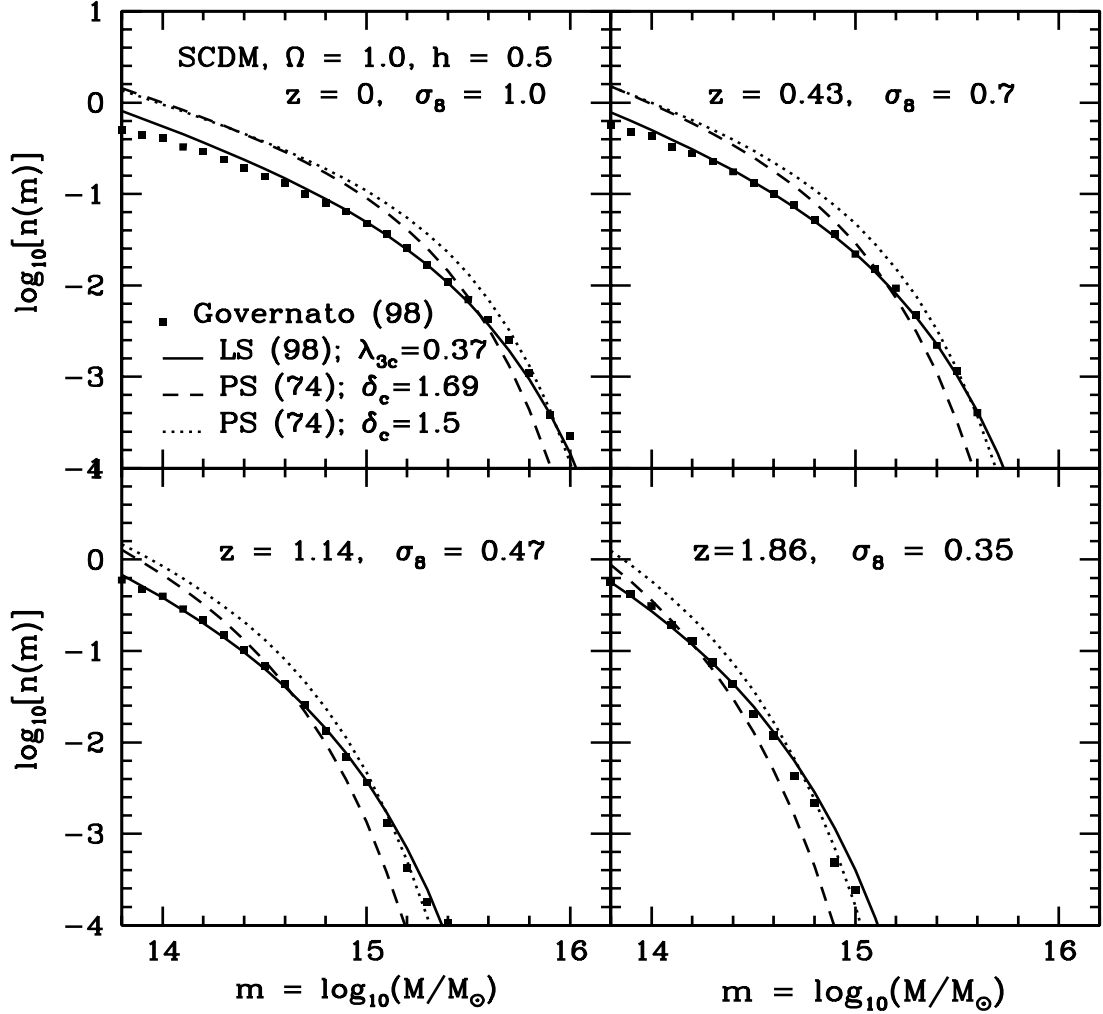,height=14.5cm,width=15.5cm}
\caption[fig2.ps]{The square dots represent the numerical data 
for the case of SCDM model with $\Omega = 1, h=0.5$. 
The solid line is our mass function with $\lambda_{3c} = 0.37$, 
and the dashed, the dotted lines
are the PS mass functions with $\delta_{c} = 1.69, 1.5$ respectively. }
\end{figure}

\end{document}